\documentclass[3p,preprint,12pt]{elsarticle}
\biboptions{sort&compress}

\usepackage{amssymb}
\usepackage{amsthm}
\usepackage{threeparttable}
\usepackage{multirow}
\usepackage{rotating}
\usepackage{lineno}

\journal{Journal of \LaTeX\ Templates}

\newcommand{\g}{$\gamma$}
\newcommand{\Lip}{$^7$Li(p,n)$^7$Be }

\begin{document}

\begin{frontmatter}

\title{Experimental $^7$Be production cross section from the \Lip reaction at E$_p = 3.5-13$ MeV}

\author[label1,label2]{\'A. T\'oth}
\ead{toth.akos@atomki.hu}

\author[label1]{Gy. Gy\"urky}

\author[label1]{E. Papp}

\author[label1]{T. Sz\"ucs\corref{cor1}}
\ead{tszucs@atomki.hu}

\affiliation[label1]{organization={Institute for Nuclear Research (Atomki)}, 
city={Debrecen}, 
country={Hungary}}

\affiliation[label2]{organization={University of Debrecen, Doctoral School of Physics}, 
city={Debrecen}, 
country={Hungary}}
\cortext[cor1]{Corresponding author.}

\begin{abstract}
The \Lip reaction is widely used as neutron source for neutron induced reaction cross section measurements, and for $^7$Be radioactive source production. 
There are two prominent structures in the excitation function, a narrow resonance between $E_\mathrm{p}= 2.2-2.3$ MeV, and a broad peak, around $E_\mathrm{p}= 5$ MeV.
There are tension between the experimental data sets both in the position and the width of this latter structure, as well as in the absolute scale of the data. 
In the present work the \Lip reaction is investigated using the activation technique, with the aim of providing comprehensive cross section data covering the second structure and connecting prior literature data sets.
The irradiations were performed with the Atomki cyclotron accelerator with pairs of thin foil targets, thus with precisely controlled reaction energy in the range of E$_\mathrm{p} = 3.5-13$ MeV. After the irradiations the activity of the samples was measured using a high-purity germanium detector. The energy uncertainty of the new data points is much smaller than in any of the previous works, while the cross section uncertainty is comparable with the most precise literature data. A consistent data set was obtained connecting the most recent and most precise literature data sets. With the new data the absolute magnitude of the \Lip reaction cross section is constrained and became more precise.
\end{abstract}


\end{frontmatter}


\section{Introduction}
\label{intro}
The \Lip proton-induced reaction is an accelerator based neutron source, which has been widely used in a large number of neutron induced cross section measurements \cite{Good58, Singh22}. It is important for technological applications \cite{Petrich08-NIMA}, and also widely used to study nuclear reactions of astrophysical interests \cite{Beer80-PRC, Wisshak06, Dillmann09, Macias20}. This reaction has become particularly interesting for studying nuclear reactions in different nucleosynthesis scenarios \cite{Ratynski88-PRC, MARTINHERNANDEZ12, Tessler16}, because it allows the neutron spectra of stars to be reproduced. In particular, the neutron spectrum corresponding to the typical temperatures in the slow neutron capture process (s-process) \cite{Kappeler11} with kT = 25 keV is widely produced by the \Lip reaction and used in neutron induced studies \cite{Heil08}. This allows the direct measurement of the so called Maxwellian averaged neutron cross sections \cite{Wisshak06-PRC} on a given isotope, which is then later used as the main parameter for model calculations to describe the s-process abundances. At higher proton energies (typically above $E_\mathrm{p} = 20$ MeV), quasi-monoenergetic (QM) neutrons can be produced for the study of neutron-induced reactions \cite{Jarosik22, Vrzalova23}.

\begin{figure}[h]
\centering
\includegraphics[width=0.5\columnwidth]{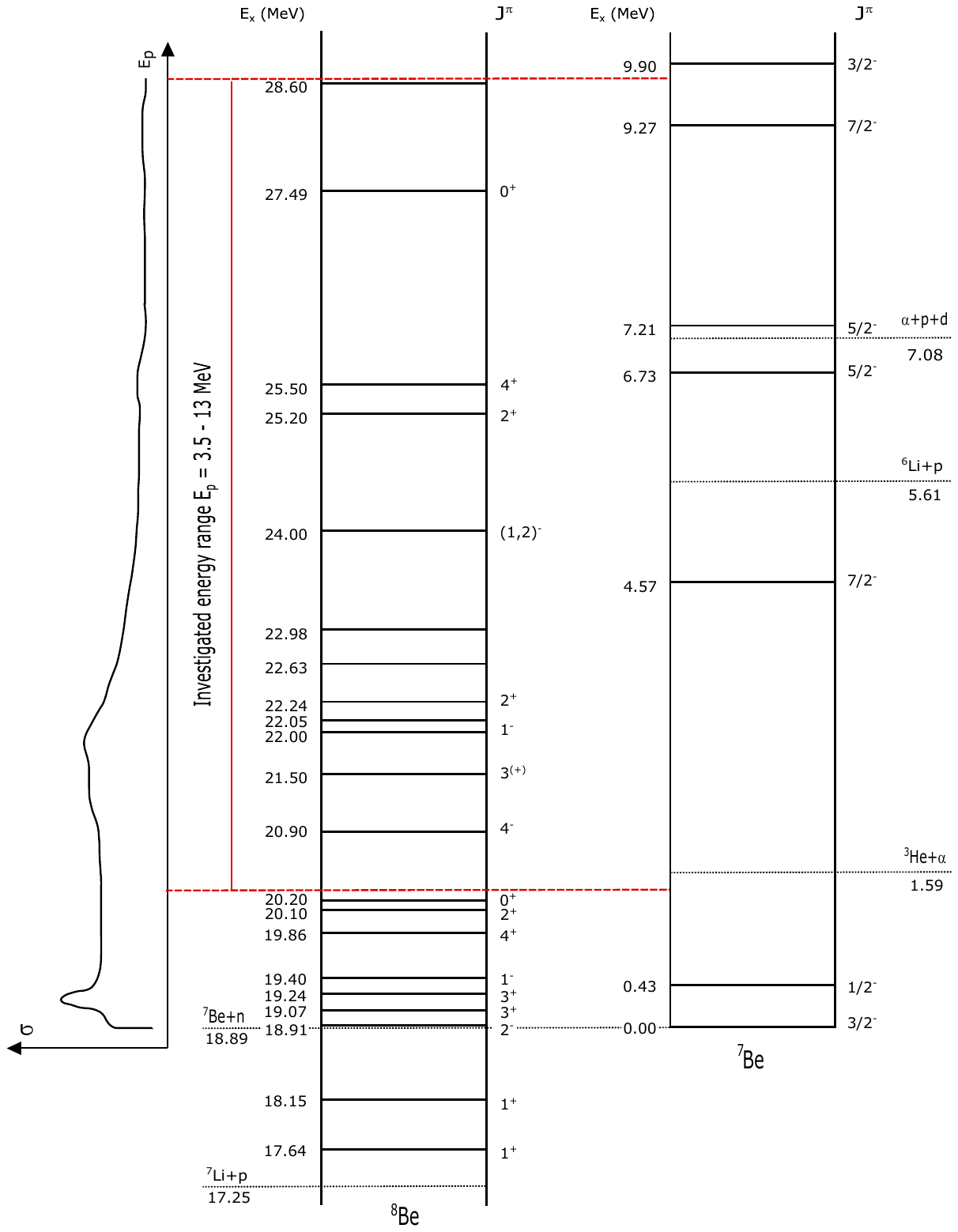}
\caption{\label{fig:levels} Partial level schemes of $^8$Be and $^7$Be \cite{Tilley02,Tilley04} aligned with the corresponding Q-value. The thresholds of several reactions are also marked, along with the energy range of the present investigation, and the schematic representation of the \Lip excitation function.}
\end{figure}

The threshold of the \Lip reaction is at $E_\mathrm{p} = 1.88$ MeV. There are two prominent peaks in the excitation function, a narrower one around $E_\mathrm{p} = 2.25$ MeV and a broad one at $E_\mathrm{p} = 5.5$ MeV. The first is a resonance corresponding to a $J^\pi=3^+$ level with $E_x = 19.235$~MeV excitation energy of the $^8$Be compound nucleus, while the second one is the effect of the sum of multiple broad levels between $E_x = 21.5 - 24.0$~MeV excitation energies. The relavant part of the $^8$Be level scheme is plotted in Fig. \ref{fig:levels}.
The reaction cross section increases steeply right above the threshold, between $E_\mathrm{p} = 1.88$ and 1.91 MeV. Several studies have been carried out in this energy region \cite{Martin16-PRC, Macklin58-PR, Taschek48-PR, Newson57-PR, Sekharan76-NIM, Gibbons59-PR}, which used neutron detection for the cross section determination. Most of these works also provided data about the first resonance mostly with consistent data sets. In this energy region, the $^7$Be reaction product is formed mainly in its ground state producing only one neutron group ($n_0$). Just above the resonance the first excited state of $^7$Be at $E_x = 0.43$~MeV can be formed leading to the emission of another neutron group ($n_1$) (see fig. \ref{fig:levels}).
Up to the second peak in the excitation function, these two neutron groups are expected to dominate the neutron yield, and only a small contribution from the three body $n + ^3\mathrm{He} + ^4\mathrm{He}$ reaction is expected \cite{Poppe76-PRC}.
In \textit{Gibbons and Macklin, 1959} \cite{Gibbons59-PR}, the total neutron production cross sections were determined. This work was the basis of the normalization of a few other works \cite{Borchers63-PR, Poppe76-PRC, Hisatake60-JPS} aimed to determine relative partial cross section and angular distributions of $n_0$ and $n_1$ production with neutron time-of-flight technique.

Just around the second structure in the \Lip excitation function, higher energy $^7$Be levels get involved in the neutron production, and an other three body channel opens ($^6$Li$+p$$+n$ beside the already opened $^3$He$+^4$He$+n$). Few works measured neutron spectra in these energy region \cite{Hisatake60-JPS, Matej2022-NIMA}. Above $E_\mathrm{p} = 20$ MeV, the $n_0$ and $n_1$ neutrons form a quasi-moenergetic (QM)  peak in the neutron spectrum. The cross section of this QM neutron production was determined in several works \cite{Simeckova21-NPA,Majerle20-EPJ,Majerle16-NPA}.
At even higher proton energies (between $E_\mathrm{p} = 60-200$ MeV) the $1/E$ dependence of the \Lip reaction cross section was investigated \cite{Ward82-PRC}.

\begin{figure}[h]
\centering
\includegraphics[width=0.85\columnwidth]{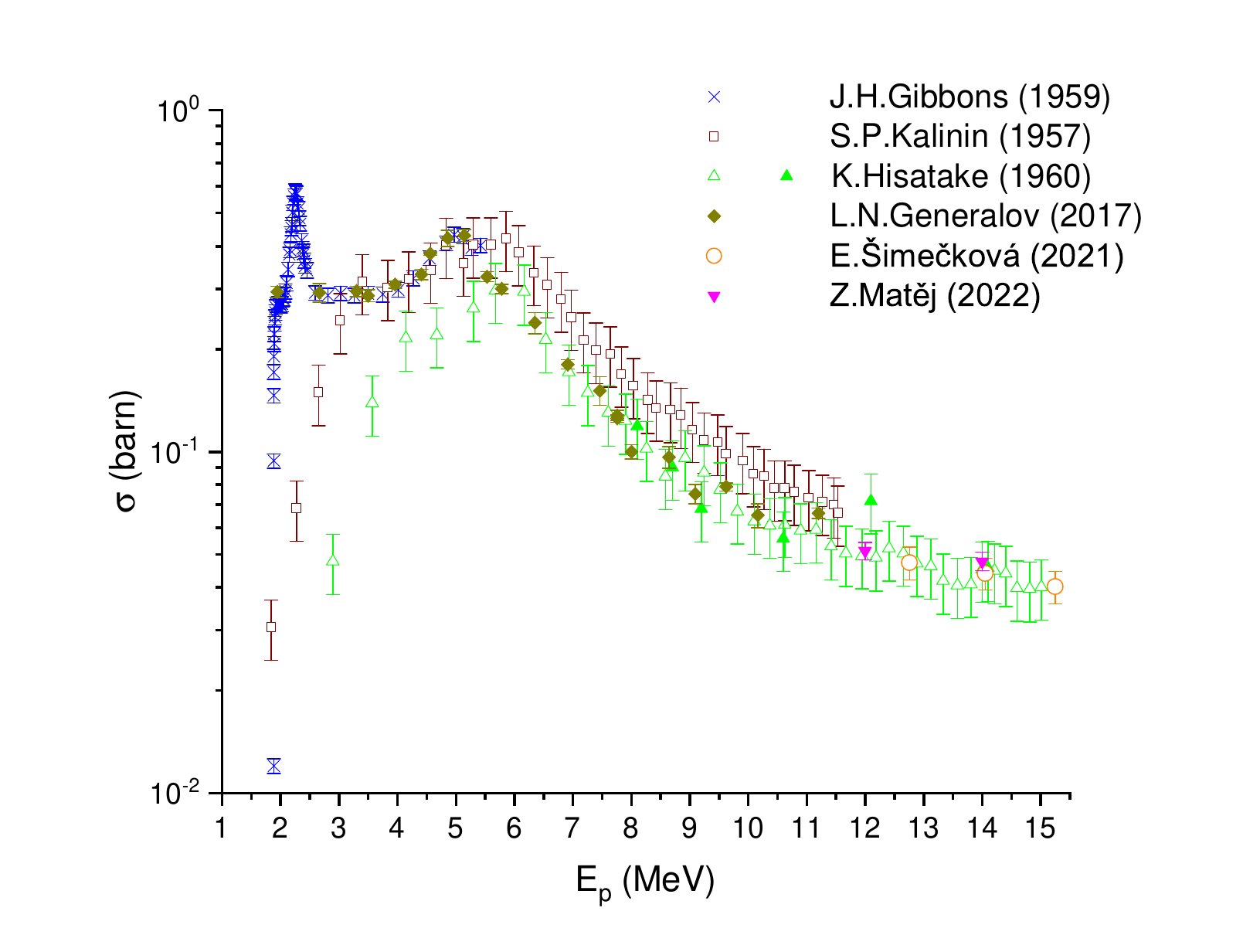}
\caption{\label{fig:literature} The literature cross section data based on $^7$Be activity determination around the second resonance \cite{Kalinin57,Hisatake60-JPS,Generalov17_BRASP,Simeckova21-NPA,Matej2022-NIMA} together with the data set measured by total neutron detection \cite{Gibbons59-PR}, which is often used for absolute normalisation for neutron detecting experiments. Data with open symbols are from stacked foils experiments  \cite{Kalinin57, Hisatake60-JPS,Simeckova21-NPA}, while those with filed symbols are from single target measurements \cite{Hisatake60-JPS, Generalov17_BRASP,Matej2022-NIMA}. The crosses indicate the neutron detection \cite{Gibbons59-PR}.}
\end{figure}

In cases, where absolute neutron production cross sections are published, those are often relative to other works, or relative to the $^7$Be production cross section. In the former case, low energy absolute measurements dominated only by the $n_0$ and $n_1$ neutron groups are used. In the latter case, the produced $^7$Be activity is measured, from which the number of reactions during the neutron detection is derived. Around the second peak in the \Lip excitation function, however, there is a tension between the previous experimental results using the activation technique \cite{Kalinin57, Hisatake60-JPS, Generalov17_BRASP} (see Fig. \ref{fig:literature}), thus the absolute normalization lacks high precision. This fact motivated the present study, where the \Lip reaction cross section was measured in the $E_\mathrm{p} = 3.5-13$ MeV energy range, at the peak structure and on its high energy tail. The well-known activation method was used employing LiF target layers on thin aluminium foil backings, while the produced $^7$Be activity was determined via $\gamma$ spectroscopy with a high purity germanium (HPGe) detector. The selected target thickness is substantially smaller than in all previous works, allowing a more precise reaction energy definition. 

This paper is organized as follows: Sec. \ref{intro} gives a short introduction, Sec. \ref{litr} summarises some selected literature works having data sets in the energy region of the present investigation. In Sec. \ref{exp} details of the target preparation, the irradiation procedure and characteristics, and the \g-ray counting are presented. The data analysis is detailed in Sec. \ref{data} and a discussion of the new results and comparison with literature data is given in Sec. \ref{results}.

\section{Data in the literature}
\label{litr}

The main purpose of the present work is to investigate the broad structure around $E_\mathrm{p} = 5$ MeV.
In this section the works providing total reaction cross section based on the determination of the activity of the $^7$Be reaction product are summarised. In those works either single targets or the stacked foil method was employed to experimentally determine the excitation function. In the former case one target is irradiated with a given beam energy, while in the latter case many targets are stacked together and exposed to the beam simultaneously. Here the beam slows in each layer, and different interaction energy is experienced by the individual targets in the stack.
The data sets are plotted in Fig. \ref{fig:literature} and summarised in Table \ref{tab:literature}.
In the table the estimated energy loss in the target foils are also presented. Those were calculated using the SRIM tables \cite{SRIM10} assuming the target thickness and composition published in the original articles.
In most of the cases, the targets were relatively thick. In such thick targets the beam particles slow significantly, since the reaction yield is the reaction cross section integrated over the energetic thickness of the targets, that may smooth out smaller structures. The effective energy attributed to a given point has to take into account the form of the cross section function within the target. In case of a rapidly varying function, this may carry substantial uncertainty, however in most of the literature works, there is no information about the effective energy determination. If the cross section function is considered constant within the target, then the effective energy is taken as the energy at the middle of the target. 
Additionally, in the particular case of stacked foils the beam energy spread increases layer-by-layer. This may also smooth out small structures in the measured excitation function.

The first report by \textit{Kalinin et al., 1957} \cite{Kalinin57} published the excitation function measured by the stacked foil method. In that work $2-10$ mg/cm$^2$ thick LiF target layers on 10 $\mu$m Al backings were used. The stack was exposed to a proton beam with an energy of $E_\mathrm{p} = 12.2 \pm 0.1$ MeV. The activity of the $^7$Be accumulated in the different layers was measured using a NaI detector and a GM-counter. In the particular case of \cite{Kalinin57}, because of the stacked foil, the beam spread became $\approx$ 30\% at the lowest energies from the initial  $\approx$ 1\%.  Additionally, in targets of these thicknesses the protons may lose up to $1.8$ MeV, which means, in the last layers the protons may slows below the threshold energy, thus the effective target thicknesses are substantially reduced. Both the effect of the beam spread and target thickness has to be also taken into account while transforming the yield data to excitation function. There is no information in the article, how this unfolding was done. It is claimed, that the first resonance is not visible because of the beam energy spread, however the effect of the target thickness is not elaborated. In the published excitation function a very broad structure is obtained around $E_\mathrm{p} = 6$ MeV.
The total uncertainty of the measured absolute cross sections was 20\%.

\textit{Hisatake et al., 1960} \cite{Hisatake60-JPS} determined the absolute \Lip reaction cross section with a few single foils, whose activity was measured by a NaI detector. Metallic lithium was used thinned to about 100 $\mu$m. The energy loss in such targets is about $0.2-0.6$ MeV. Besides the single foils a more complete excitation function was measured with stacked foils, where the absolute scale was normalized to the single foil data. The activity of the foils from the stack was determined with an other NaI detector in well geometry. No details are provided about the normalization and some of the single target data differs significantly from the stacked foil points. There is no information either how the effect of the target thickness was taken into account, or if any backing was applied. Without a backing substantial amount of $^7$Be may escape from the target because of the reaction kinematics. Considering the published single target energies, the calculated $^7$Be recoil energies and the angular distribution, up to 15\% of the activity may be lost. In case of the stack, part of the activity from a given foil may be accumulated at the subsequent target, altering the slope of the measured excitation function.
The paper also claimed better than 20\% precision of the data, however about 50\% difference is seen in the absolute scale compared to \cite{Kalinin57}.

In a recent study by \textit{Generalov et al., 2017} \cite{Generalov17_BRASP}, the reaction was revisited using single targets. The cross sections of several reactions including the \Lip were investigated using the activation technique. The irradiations were performed between $E_\mathrm{p}= 2-10$ MeV beam energies. As targets lithium fluoride was evaporated onto copper or tantalum backing with a thickness between $0.6-1$ mg/cm$^2$. The energy loss in the targets is presented in the article, so are the effective beam energies, however no information is given about the derivation and uncertainty of those values. The targets were separated from the beamline vacuum by a 15 $\mu$m aluminium window and a helium stream during the irradiations. There is no information in the article about the thickness precision of the exit window, which causes already some energy loss ($0.12-0.4$ MeV). The \g -ray yield was measured with two germanium detectors.
The uncertainty of the measured cross sections varies between 1\% and 10\%. 

In \textit{Simeckova et al., 2021} \cite{Simeckova21-NPA} the stacked foil technique was used again. The energy range of this experiment ($E_\mathrm{p}= 12.5 - 34$ MeV) slightly overlaps with that of the present study. Stack of Li foils with about 500 $\mu$m thickness separated by tantalum degrader and copper monitor foils were irradiated. The energy loss in a given foil is about 0.9 MeV in the overlapping energy range with the present work.
The activity of the samples was measured with HPGe detectors. The mean beam energy in a given sample was determined by matching the experimental excitation function of the copper monitors to its recommended literature reference. From this procedure the claimed energy uncertainty is 3\% for each of the data points. There is no information how the effect of the target thickness, thus the effective reaction energy is calculated, however in this energy region, the cross section is flat causing only a minor uncertainty in the slope or absolute value of the excitation function. Similarly to the previous stacked foil work using metallic lithium, there is no mention of any backing foils, thus $^7$Be may have escaped from the targets. Because of the thicker target, this is about 3\% below 15 MeV reaction energy well within the claimed total uncertainty of $10-11$\%.

In the most recent work by \textit{Matej et al., 2022} \cite{Matej2022-NIMA}, the \Lip reaction cross section was determined using lithium metal targets of 500 $\mu$m thickness at proton energies of 12.4 and 14.4 MeV. The proton energy loss in the targets is 0.9 MeV and 0.8 MeV, respectively. 
In Table 8 of the original article, the initial beam energies are qoued as the reaction energy at the center of the target, however the proton energies at the middle of the target shall be $E_\mathrm{p}=12.0$ MeV and $E_\mathrm{p}=14.0$ MeV, respectively. These values are found in the EXFOR \cite{Exfor14} data table D1028003 referenced as provided by the authors, thus we are using the EXFOR values. No backing foil was used, thus $^7$Be can escape from the foil. The range of $^7$Be in lithium with their average energies calculated from the reaction kinematic is about $13-14$ $\mu$m, which is about 3\% of the total target thickness. The uncertainties of the cross section values are 5.9\% and 6.3\%, respectively. 

\begin{table}[ht]
\caption{The relevant literature datasets using the activation method, listed from the Experimental Nuclear Reaction Data Library (EXFOR) database \cite{Exfor14}. For the derivation of the listed quantities, see text.}
\label{tab:literature}
\centering
\begin{threeparttable}
\begin{tabular}{ | c | c | c | c | c | c | c | c | }	
\hline					
							&		& Proton 		& 				& Energy loss	& Cross section \\
 Ref.						& Year	& energy range	& Targets		& in the target	& uncertainty \\
 							&		& (MeV)			& 				& (MeV)			&  \\ 						

\hline \hline

\cite{Kalinin57} 				& 1957	& $1.84 - 11.5$ 	& stacked LiF	& $0.1 - 1.8$ 		& 20\% \\
\hline
\cite{Hisatake60-JPS}		& 1960	& $2.9 - 15$ 	& single/stacked Li foils & $0.2 - 0.6$ 	& 20\% \\
\hline
\cite{Generalov17_BRASP} 	& 2017 & 1.95 - 10.2		& single LiF		& $0.01 - 0.1$\tnote{a}  	& $1-10$\% \\
\hline
\cite{Simeckova21-NPA} 		& 2021 & 12.8 - 34		& stacked Li foils	& $0.6 - 0.8$ 	& $10-11$\% \\
\hline
\cite{Matej2022-NIMA} 		& 2022 & 12 - 14		& single Li foils	& $0.8 - 0.9$ 	& $5.9-6.3$\% \\
\hline
\multicolumn{2}{|c|}{This work}	& 3.5 - 13			& pairs\tnote{b}  ~of single LiF& $0.003 - 0.008$  & $3.9-5.3$\% \\
\hline
\end{tabular}
\begin{tablenotes}
\item[a]{As presented in the paper.}
\item[b]{See sec. \ref{irrad} for details.}
\end{tablenotes}
\end{threeparttable}
\end{table}

\section{Experimental details}
\label{exp}

\subsection{Targets preparation}
\label{target}

Lithium fluoride with natural isotopic composition (7.59\% for $^6$Li and 92.41\% for $^7$Li, respectively) was evaporated onto high purity, 10~${\mu}m$ nominal thickness aluminium backings. This is a minimum necessary thickness to stops the  $^7$Be nuclei. Taking into account the reaction kinematic at the maximum beam energy (E$_p = 13$ MeV) the $^7$Be recoil have a maximum of 5.31 MeV kinetic energy, thus a range of 7.4 ${\mu}m$ in aluminium according the SRIM tables \cite{SRIM10}. 

For the target production, a Leybold Univex 350 vacuum evaporator was used. The LiF was melted in a molybdenum boat heated by electric current. The melting point of the boat material is more than 1700 $^\circ$C higher than that of the LiF.
The target material is deposited on the backing in a circular surface of 12 mm in diameter, which is defined by the size of the hole on the foil holder. The boat-backing distance was chosen to be 12 cm to create a suitable homogeneous layer. At this distance, the deposition inhomogeneity caused by the distance difference of the center and the edge of the target from the boat is below 0.25\%.
The resulting LiF layer thickness was determined by weighing. The mass of the backing foils was measured before and after the evaporation. The difference of the two mass measurements was used to determine the mass of the evaporated material. The areal number density of the active target atoms was calculated assuming standard 1:1 Li:F stoichiometry of the compound.
The areal density was between $0.08-0.12$ mg/cm$^2$, which is summarized in Table \ref{tab:targets}. The uncertainties of the LiF areal density varied between 1.8\% and 4.1\%, which is dominated by the uncertainty of the weighing. Each sample was weighed five times and the standard deviation of the resulting weights was used in the uncertainty calculation.

\begin{sidewaystable}
\caption{The proton energies at the entrance surface of the front and rear targets. One row corresponds to one irradiation.  The thicknesses of targets, backing foils and the degrader foils are also presented along with the accumulated charge in a given assembly. For the $\#3-\#4$ and $\#5-\#6$ irradiations no degrader was used.}
\label{tab:targets}
\centering
\begin{tabular}{ | c | c | c | c | c | c | c | c | c |}	
\hline					
Target	&	 \multirow{2}{*}{$E_\mathrm{p}$} 	&	LiF areal	& Al	backing	&	degrader foil	&	Target	& \multirow{2}{*}{$E_\mathrm{p}$'}	&	LiF areal	& Integrated	 \\
 	 num.	&   	 & density	&	thickness & thickness	&	num.	&		&	density	&	charge	  \\
 {[front]}	&	(MeV)	&	(mg/cm$^2$)	&	($\mu$m)	&	($\mu$m)	&	{[rear]} &	(MeV)	&	(mg/cm$^2$)	&	(mC)		\\
\hline				
\#1	&	4.000\,$\pm$\,0.012	&	0.105\,$\pm$\,0.003	&	10.01\,$\pm$\,0.08	&	10.58\,$\pm$\,0.05 (Al)	&	\#2	&	3.600\,$\pm$\,0.018	&	0.111\,$\pm$\,0.003	&	2.26\,$\pm$\,0.07	\\
\#3	&	4.500\,$\pm$\,0.014	&	0.121\,$\pm$\,0.004	&	9.93\,$\pm$\,0.08	&	-	&	\#4	&	4.322\,$\pm$\,0.016	&	0.089\,$\pm$\,0.003	&	2.07\,$\pm$\,0.06	\\
\#5	&	5.000\,$\pm$\,0.015	&	0.091\,$\pm$\,0.003	&	10.10\,$\pm$\,0.08	&	-	&	\#6	&	4.835\,$\pm$\,0.017	&	0.096\,$\pm$\,0.004	&	2.43\,$\pm$\,0.07	\\
\#7	&	5.500\,$\pm$\,0.017	&	0.110\,$\pm$\,0.003	&	10.01\,$\pm$\,0.08	&	10.58\,$\pm$\,0.05 (Al)	&	\#8	&	5.189\,$\pm$\,0.019	&	0.112\,$\pm$\,0.004	&	1.70\,$\pm$\,0.05	\\
\#9	&	6.000\,$\pm$\,0.018	&	0.097\,$\pm$\,0.004	&	10.00\,$\pm$\,0.07	&	10.58\,$\pm$\,0.05 (Al)	&	\#10	&	5.710\,$\pm$\,0.020	&	0.085\,$\pm$\,0.002	&	2.49\,$\pm$\,0.07	\\
\#11	&	7.000\,$\pm$\,0.021	&	0.096\,$\pm$\,0.002	&	10.12\,$\pm$\,0.08	&	31.86\,$\pm$\,0.26 (Al)	&	\#12	&	6.473\,$\pm$\,0.029	&	0.104\,$\pm$\,0.002	&	2.49\,$\pm$\,0.07	\\
\#13	&	8.000\,$\pm$\,0.024	&	0.109\,$\pm$\,0.004	&	9.95\,$\pm$\,0.09	&	31.86\,$\pm$\,0.26 (Al)	&	\#14	&	7.528\,$\pm$\,0.030	&	0.109\,$\pm$\,0.003	&	2.38\,$\pm$\,0.07	\\
\#15	&	9.000\,$\pm$\,0.027	&	0.109\,$\pm$\,0.002	&	9.91\,$\pm$\,0.08	&	31.86\,$\pm$\,0.26 (Al)	&	\#16	&	8.571\,$\pm$\,0.031	&	0.102\,$\pm$\,0.002	&	2.61\,$\pm$\,0.08	\\
\#17	&	10.000\,$\pm$\,0.030	&	0.117\,$\pm$\,0.003	&	10.07\,$\pm$\,0.08	&	15.16\,$\pm$\,0.15 (Ta)	&	\#18	&	9.422\,$\pm$\,0.035	&	0.098\,$\pm$\,0.003	&	2.35\,$\pm$\,0.07	\\
\#19	&	11.000\,$\pm$\,0.033	&	0.092\,$\pm$\,0.004	&	10.13\,$\pm$\,0.08	&	15.16\,$\pm$\,0.15 (Ta)	&	\#20	&	10.460\,$\pm$\,0.037	&	0.091\,$\pm$\,0.003	&	1.15\,$\pm$\,0.03	\\
\#21	&	12.000\,$\pm$\,0.036	&	0.096\,$\pm$\,0.003	&	10.17\,$\pm$\,0.04	&	20.64\,$\pm$\,0.21 (Ta)	&	\#22	&	11.337\,$\pm$\,0.042	&	0.101\,$\pm$\,0.002	&	1.25\,$\pm$\,0.04	\\
\#23	&	13.000\,$\pm$\,0.039	&	0.097\,$\pm$\,0.003	&	10.15\,$\pm$\,0.08	&	20.64\,$\pm$\,0.21 (Ta)	&	\#24	&	12.374\,$\pm$\,0.044	&	0.079\,$\pm$\,0.003	&	1.24\,$\pm$\,0.04	\\
\hline
\end{tabular}
\end{sidewaystable}

\subsection{Irradiation}
\label{irrad}

\begin{figure}[h]
\centering
\includegraphics[width=0.8\columnwidth]{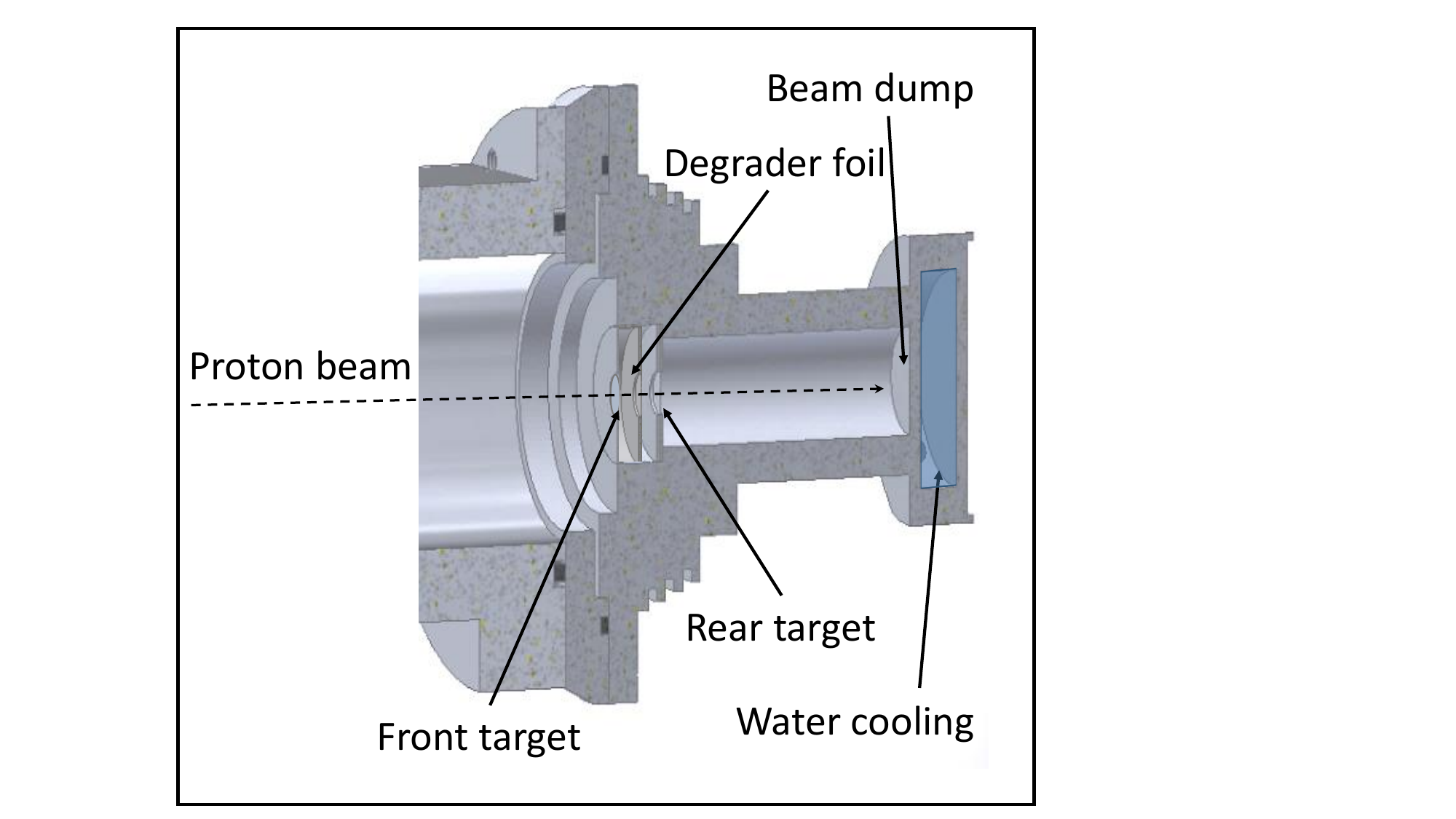}
\caption{\label{fig:setup} The target assembly.}
\end{figure}

A thin target configuration was used, where the beam energy loss in the target is only a small fraction of the initial beam energy.
This allowed the irradiation of two samples placed behind each other at the same time.
The layout of the target assembly is shown in Fig. \ref{fig:setup}. 

The proton beam was provided by the MGC-20 cyclotron accelerator of Atomki. The \Lip reaction was investigated in the following steps: in the $E_\mathrm{p} = 3.5-6$ MeV range, $\approx$ 0.25 MeV intervals, while in the E$_p = 6-13$ MeV range, $\approx$ 0.5 MeV intervals.
The activation chamber acted as a Faraday cup. A voltage of $-300$ V was applied at the entrance of the chamber to eliminate the effect of any secondary electrons that may be generated either in the 5-mm diameter beam defining aperture or in the targets falsifying the current reading. These allow the determination of the number of bombarding particles via charge measurement with a 3\% accuracy. The unreacted beam was absorbed in a water cooled 0.5 mm thick tantalum disc behind the target assembly. 
Details about the irradiations, employed targets and degraders can be found in Table \ref{tab:targets}. 

The front targets experienced the proton beam with its initial energy and 0.3\% energy uncertainty of the cyclotron and beam transport system.
The total energy loss in the LiF layer is maximum 8 keV for the lowest beam energy, and 3 keV for the highest beam energy. Even if this is almost negligible, the effective energy is taken at the center of the target layer. 

The exact thickness of the Al backing needed for the effective beam energy determination at the rear foils were measured with two methods. First, the surface area of the square shaped foils was measured in a microscope. Then from its weight assuming standard aluminium density the thickness was determined. Second, the thickness of the Al backing foils was also determined by alpha energy loss measurement. The energy loss of alpha particles from a triple-isotope alpha source was determined in an ORTEC Soloist alpha spectrometer. The two methods gave consistent results, and their average was used in the data analysis.
For most of the irradiations, the energy loss in the backing was not enough to reach the desired proton energy at the rear target. In these cases energy degrader foils between the two targets were utilized. Either several layers of the above mentioned 10 ${\mu}m$ nominal thickness aluminium, or for the highest beam energies tantalum foils were used. The thickness of the aluminium degraders were determined similarly to that of the backings, while the tantalum thicknesses were determined from their weight and surface area only.
The beam energy at the rear target was determined using SRIM \cite{SRIM10} simulations to give an accurate estimate of the energy loss in the front target, in its backing foil and in the degrader layers (if any). These simulations also show negligible effect of backscattering of the beam particles (below 0.01\%).
At the rear targets the energy uncertainty is the squared sum of the following partial uncertainties. Initial beam energy uncertainty (0.3\%), energy loss uncertainty in the backing and degrader foils including the thickness uncertainty of those and the uncertainty of the proton stopping power (altogether $0.1-0.3\%$ compared to the beam energy). 

The length of the irradiations was relatively short, between 15-25 minutes while the beam current was 2 ${\mu}A$ except for energies above $E_\mathrm{p} = 10$ MeV, where the beam current was 1 ${\mu}A$ to reduce the undesirable activity of parasitic nuclei with short half-lives produced on the impurities of the backings.
The maximum power deposition in the backing foils was 0.4~W, which may cause a temperature increase of 35 $^\circ$C at the middle of the targets with respect to the edges being in good heat contact with the water-cooled chamber.
The resulting foil temperature was well below the melting point of LiF (about 850°C) or the Al backing (660°C), thus no material loss is expected during the irradiation. This was also checked by measuring the weight of selected targets after the irradiation, and mass loss was not observed.   

\subsection{\g-counting}
\label{count}

The $^7$Li+p reaction produces a compound nucleus, $^8$Be, in a highly excited state. The $^8$Be is transformed into the $^7$Be nucleus following the emission of a neutron. The $^7$Be nuclei are implanted in the Al backing foil during the irradiations. The $^7$Be beta decays (with a half-life of 53.22 days) with a probability of 10.44\% \cite{Tilley02} to $^7$Li first excited state. Upon de-excitation, a \g  -photon with energy of E$_\gamma=477.6$ keV \cite{Tilley02} is emitted, which was detected off-line, after the irradiations with a HPGe detector.

\begin{figure}[h]
\centering
\includegraphics[width=0.85\columnwidth]{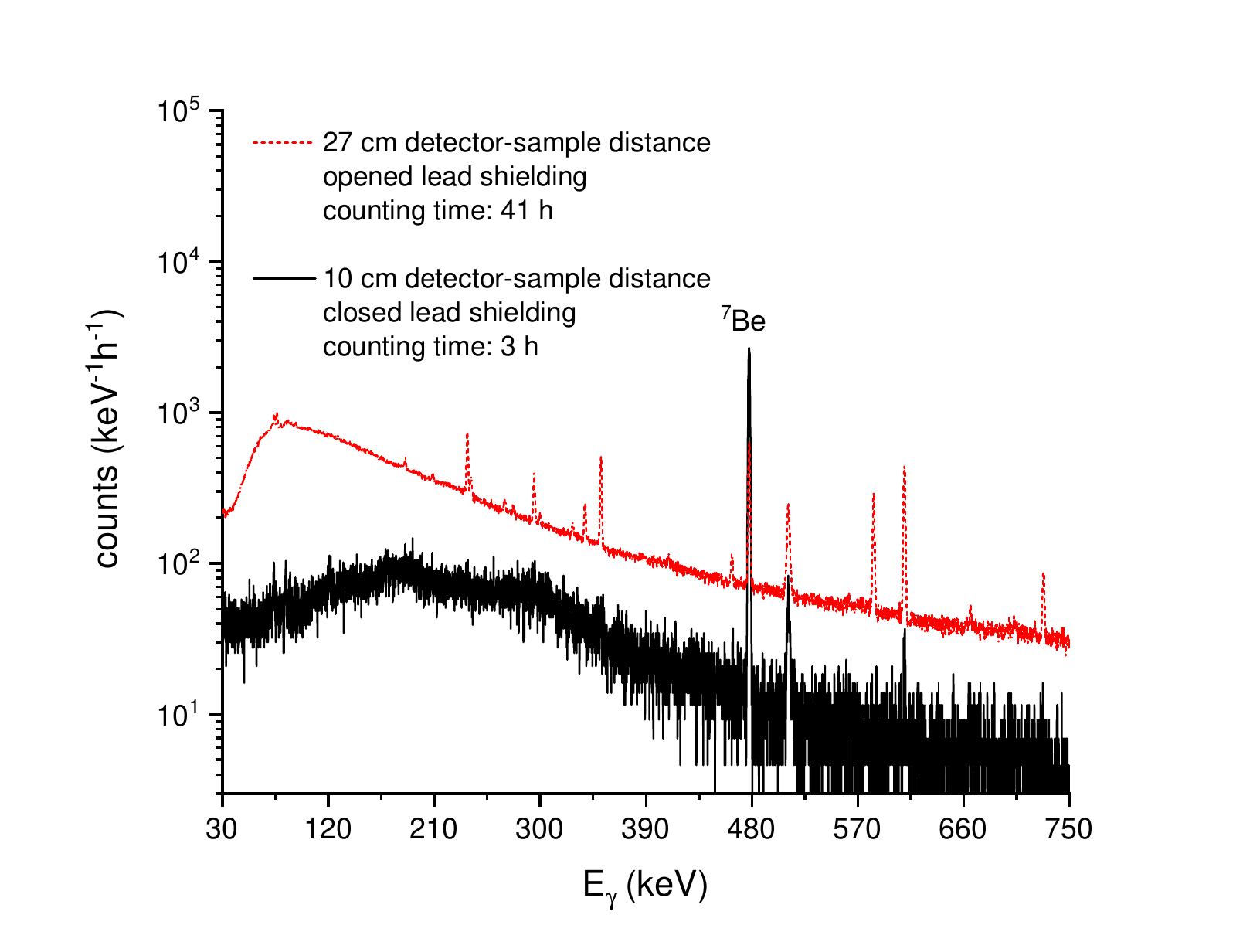}
\caption{\label{fig:spectrum} Comparison of the gamma spectra measured at 10 cm and 27 cm detector-sample distance. At a distance of 10 cm, background reduction effect of the fully closed lead shielding is clearly visible.}
\end{figure}

The efficiency of HPGe detector was determined using point-like calibrated sources of known activity ($^{152}$Eu and $^{133}$Ba). To determine the efficiency, high intensity transitions were used from both sources. In far geometry, where the true coincidence summing effect \cite{Gilmore08} is negligible, at a source and detector distance of 27 cm, the efficiency was determined with an uncertainty of less than 1\%.
The actual samples have the activity distributed within the 5-mm diameter beam spot, which can also be considered point-like on a 0.01\% level, well within the quoted efficiency uncertainty.
Due to the large number of irradiated samples, \g-countings were performed in 10 cm geometry. This reduced significantly the counting time and due to the fully enclosed lead shielding, the background in the measured spectra was an order of magnitude smaller, see in Fig. \ref{fig:spectrum}. An efficiency ratio between the two counting geometry was determined with one of the stronger $^7$Be sources. Based on this scaling factor and the uncertainty of the far geometry efficiency, the efficiency is known with 1.5\% accuracy in the 10 cm geometry.

\begin{figure}[h]
\centering
\includegraphics[width=0.85\columnwidth]{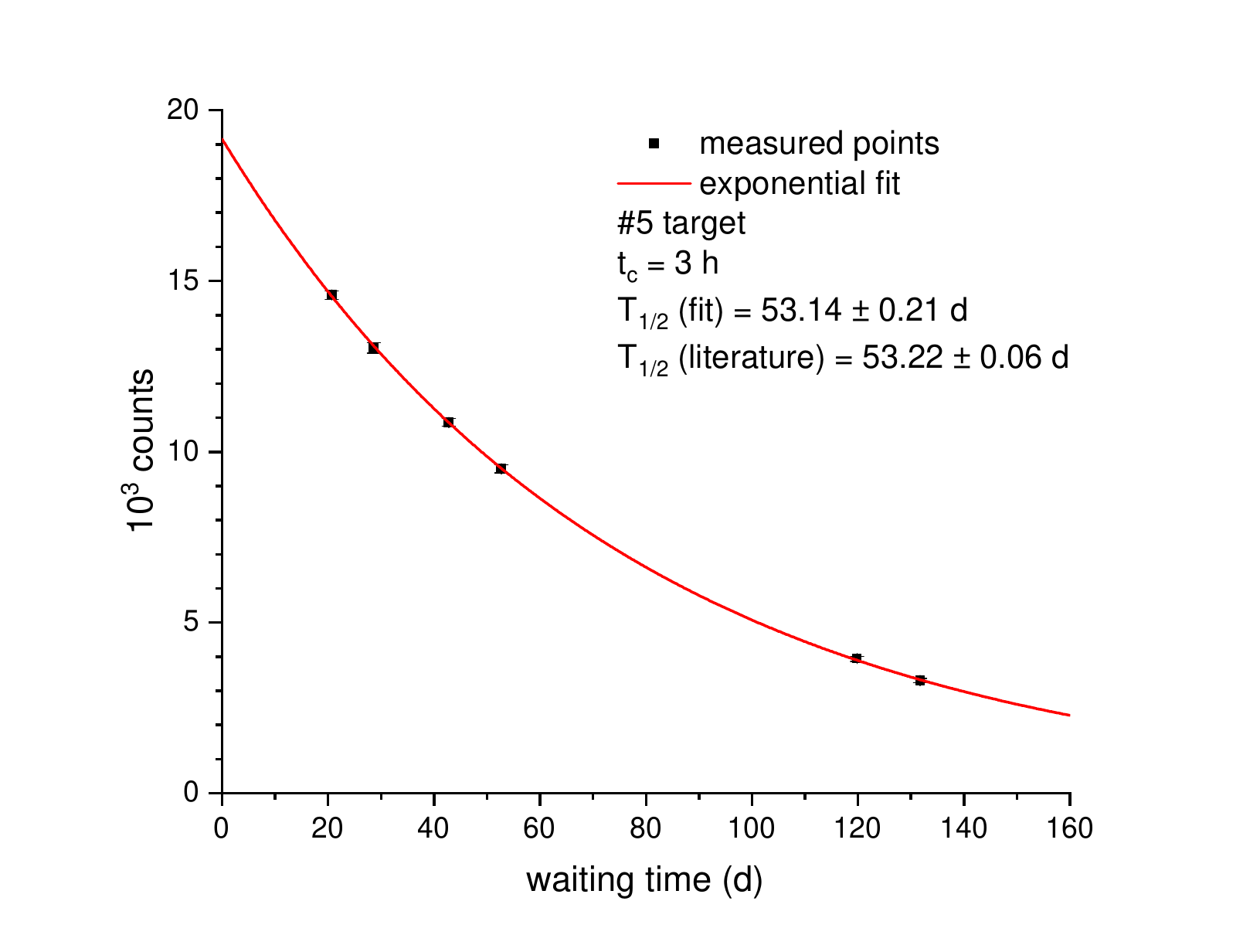}
\caption{\label{fig:decay} Decay curve fitted from six countings of $\#5$ target. The length of the countings were 3 h in each case. The literature half-life, the half-life from the exponential fit and their uncertainties are shown.}
\end{figure}

The irradiated samples were placed in front of the detector, each sample at least at two different times to follow the decay of $^7$Be, and check for parasitic activities which may spoil the peak of interest. The peak count rate difference between the two counting was consistent for each target taking into account the $^7$Be half-life. One target was measured at six different times to investigate the decay curve. In this case, the decay of $^7$Be was followed for slightly more than two half-lives, as shown in Fig \ref{fig:decay}. The half-life value obtained from the fit is in good agreement with its literature value.
The targets between $E_\mathrm{p} = 3.5-7$ MeV were usually in front of the detector for $5-6$ hours. Occasionally, the second counting was longer ($24-25$ hours). Between E$_p = 7-13$ MeV, the counting time was $15-20$ hours, in some cases $50-60$ hours for the second counting to reach the desired statistics.

\subsection{Data analysis}
\label{data}
The cross section ($\sigma$) of the \Lip reaction was determined using the following equation:

\begin{equation}
\sigma = A \left(N_t \frac{C}{e} \epsilon_\gamma\eta e^{-\lambda t_w} (1-e^{-\lambda t_c}) \right)^{-1},
\end{equation}

where $A$ is the net peak area determined by fitting the gamma spectrum using a lognormal function and a linear background. $N_t$ is the areal number density of active target atoms, $C/e$ is the  number of impinged protons ($C$ as the integrated charge, and $e$ as the elementary charge). Since the half-life of the reaction product is much longer than the irradiation time, the number of decays during the irradiation is negligible.
$\epsilon_\gamma$ is the branching ratio of the emitted \g-line, while $\eta$ is the detection efficiency. $\lambda$ is the decay constant of $^7$Be, while $t_w$ and $t_c$ are the waiting time between the end of irradiation and the start of gamma counting and the length of the gamma counting, respectively.

\section{Results and conclusions}
\label{results}

The new cross section data of the \Lip reaction are summarised in Table \ref{tab:run}, where the values are given with the total uncertainty ($3.9-5.3$\%) along with the effective proton energies and their uncertainties.
\begin{table}[h]
\caption{The effective beam energies and the cross sections of the \Lip reaction.}
\label{tab:run}
\centering
\begin{tabular}{ | c | c | c | }	
\hline				
Target 	&	$ E_{eff,p} \pm \Delta E_{eff,p} $	&  		$\sigma \pm \Delta \sigma_{total}$	  \\
 	number		& (MeV)	& (barn)  \\	
\hline
\#1		& 	3.596\,\,$\pm$\,\,0.018	&	0.2848\,$\pm$\,0.0123	\\
\#2		&	3.996\,$\pm$\,0.012	& 	0.3060\,$\pm$\,0.0143	\\	
\#3		&	4.319\,$\pm$\,0.016	&	0.3308\,$\pm$\,0.0171	\\				
\#4		&	4.496\,$\pm$\,0.014	&	0.3389\,$\pm$\,0.0155	\\
\#5		&	4.832\,$\pm$\,0.017	&	0.3894\,$\pm$\,0.0205	\\						
\#6		&	4.997\,$\pm$\,0.014	&	0.3973\,$\pm$\,0.0177	\\ 						
\#7		&	5.185\,$\pm$\,0.019	&	0.3938\,$\pm$\,0.0190	\\						
\#8		&	5.497\,$\pm$\,0.017	&	0.3810\,$\pm$\,0.0162	\\
\#9		&	5.708\,$\pm$\,0.020	&	0.3106\,$\pm$\,0.0138	\\
\#10	&	5.997\,$\pm$\,0.018	&	0.2781\,$\pm$\,0.0139	\\	
\#11	&	6.471\,$\pm$\,0.029	&	0.2142\,$\pm$\,0.0083	\\
\#12	&	6.998\,$\pm$\,0.021	&	0.1676\,$\pm$\,0.0065	\\		
\#13	&	7.526\,$\pm$\,0.030	&	0.1273\,$\pm$\,0.0054	\\
\#14	&	7.998\,$\pm$\,0.024	&	0.1098\,$\pm$\,0.0052	\\	
\#15	&	8.569\,$\pm$\,0.031	&	0.0888\,$\pm$\,0.0034	\\				
\#16	&	8.998\,$\pm$\,0.027	&	0.0878\,$\pm$\,0.0036	\\
\#17	&	9.420\,$\pm$\,0.035	&	0.0791\,$\pm$\,0.0035	\\						
\#18	&	9.998\,$\pm$\,0.030	&	0.0685\,$\pm$\,0.0030	\\ 						
\#19	&	10.459\,$\pm$\,0.037	&	0.0627\,$\pm$\,0.0029	\\						
\#20	&	10.998\,$\pm$\,0.033	&	0.0594\,$\pm$\,0.0032	\\
\#21	&	11.335\,$\pm$\,0.042	&	0.0623\,$\pm$\,0.0025	\\
\#22	&	11.998\,$\pm$\,0.036	&	0.0564\,$\pm$\,0.0025	\\	
\#23	&	12.373\,$\pm$\,0.044	&	0.0595\,$\pm$\,0.0031	\\
\#24	&	12.999\,$\pm$\,0.039	&	0.0504\,$\pm$\,0.0024	\\	
\hline		
\end{tabular}
\end{table}
The shown total uncertainty is the quadratic sum of the following statistical and systematic uncertainties.
The uncertainty of the peak area determination was between $0.3-1\%$, while the uncertainty of the number of target atoms was $1.8-4.1\%$, quadratically added together as statistical uncertainty.
The systematic uncertainty, affecting the scale of the complete dataset is 3.4\%, as the quadratic sum of the charge integration (3\%), $\gamma$ branching ratio (0.4\%), detection efficiency (1.5\%).

\begin{figure}[h]
\centering
\includegraphics[width=0.85\columnwidth]{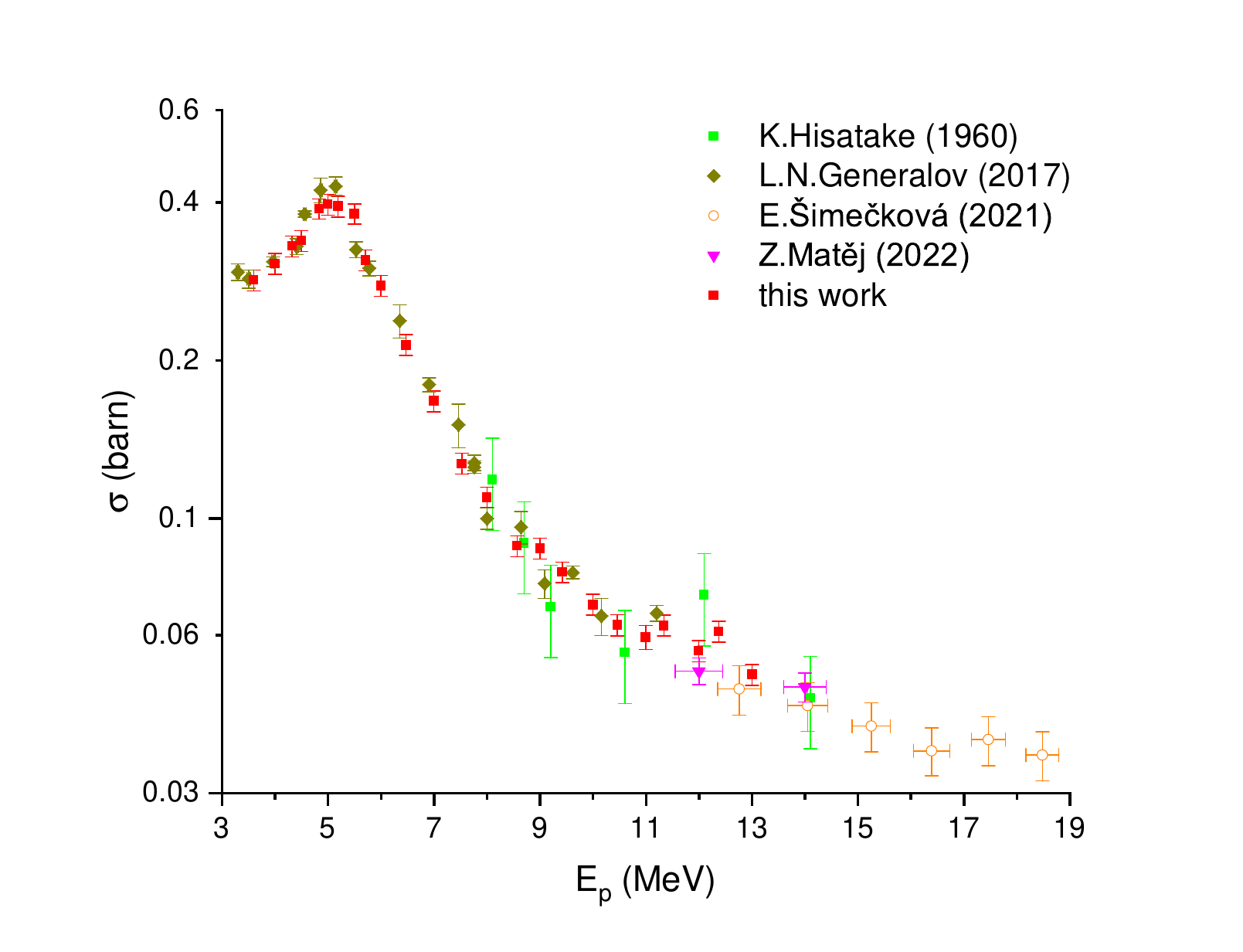}
\caption{\label{fig:results} The new cross section results compared with other studies \cite{Hisatake60-JPS,Generalov17_BRASP,Simeckova21-NPA,Matej2022-NIMA}. In case of \cite{Simeckova21-NPA,Matej2022-NIMA} the horizontal error bars are the respective energetic target thicknesses, in the other cases those are smaller than the size of the points.}
\end{figure}

In Fig. \ref{fig:results}, the experimental data are plotted and compared to the data from previous papers.
The new results are in good agreement with \cite{Generalov17_BRASP}, but in terms of energy uncertainty, more accurate data have been obtained. Furthermore, the higher energy points overlap with the \cite{Simeckova21-NPA} and \citep{Matej2022-NIMA} data series showing consistent cross sections. 
A hint of fine structure is seen in the new dataset, especially the points at 11.3 and 12.4 MeV are higher than the general trend. The highest energy data point of \cite{Generalov17_BRASP} also shows the up-bend at 11.2 MeV, while the 12.1 MeV point of \cite{Hisatake60-JPS} is also higher than the trend. \cite{Simeckova21-NPA,Matej2022-NIMA} used much thicker targets, thus the obtained cross section is an average in the marked energy regions, thus missing the possible fine structure.
A study with much smaller energy steps would unravel the fine structure, if any, but that is out of the scope of the present paper.

\section{Summary}
\label{summary}

The \Lip reaction cross section was experimentally determined in the energy range of E$_\mathrm{p} = 3.5-13$ MeV with the activation method. The thin targets allowed to scan the excitation function with a better energy determination than in previous works. The experimental uncertainties are kept under control, resulting in an overall accuracy of 5\% of the cross section. The new results are consistent with previous works, and have higher precision both in the magnitude and the energy of the data points. 
  
\section*{ACKNOWLEDGEMENTS}
This work was supported by \mbox{NKFIH} (OTKA FK134845 and K134197), \linebreak by \mbox{GINOP-2.3.3-15-2016-00029}, and by the New National Excellence Programs of the Ministry of Human Capacities of Hungary under nr. \mbox{\'UNKP-22-3-II-DE-31}. T.S. acknowledges support from the J\'anos Bolyai research fellowship of the Hungarian Academy of Sciences.


\begin{thebibliography}{10}
\expandafter\ifx\csname url\endcsname\relax
  \def\url#1{\texttt{#1}}\fi
\expandafter\ifx\csname urlprefix\endcsname\relax\def\urlprefix{URL }\fi
\expandafter\ifx\csname href\endcsname\relax
  \def\href#1#2{#2} \def\path#1{#1}\fi

\bibitem{Good58}
W.~M. {Good}, J.~H. {Neiler}, J.~H. {Gibbons}, {Neutron Total Cross Sections in
  the kev Region by Fast Time-of-Flight Measurements}, Physical Review 109~(3)
  (1958) 926--933.
\newblock \href {https://doi.org/10.1103/PhysRev.109.926}
  {\path{doi:10.1103/PhysRev.109.926}}.

\bibitem{Singh22}
R.~K. {Singh}, N.~L. {Singh}, R.~D. {Chauhan}, M.~{Mehta}, S.~V.
  {Suryanarayana}, R.~{Makwana}, B.~K. {Nayak}, H.~{Naik}, T.~N. {Nag},
  K.~{Katovsky}, {Systematic study of the (n, 2n) reaction cross section for
  $^{121}$Sb and $^{123}$Sb isotopes}, Chinese Physics C 46~(5) (2022) 054002.
\newblock \href {https://doi.org/10.1088/1674-1137/ac4a5a}
  {\path{doi:10.1088/1674-1137/ac4a5a}}.

\bibitem{Petrich08-NIMA}
D.~{Petrich}, M.~{Heil}, F.~{K{\"a}ppeler}, J.~{Kaltenbaek}, E.~P. {Knaetsch},
  K.~{Litfin}, D.~{Roller}, W.~{Seith}, R.~{Stieglitz}, F.~{Voss}, S.~{Walter},
  {A neutron production target for FRANZ}, Nuclear Instruments and Methods in
  Physics Research A 596~(3) (2008) 269--275.
\newblock \href {https://doi.org/10.1016/j.nima.2008.08.131}
  {\path{doi:10.1016/j.nima.2008.08.131}}.

\bibitem{Beer80-PRC}
H.~{Beer}, F.~{K{\"a}ppeler}, {Neutron capture cross sections on $^{138}$Ba,
  $^{140,142}$Ce, $^{175,176}$Lu, and $^{181}$Ta at 30 keV:Prerequisite for
  investigation of the $^{176}$Lu cosmic clock}, Phys. Rev. C 21~(2) (1980)
  534--544.
\newblock \href {https://doi.org/10.1103/PhysRevC.21.534}
  {\path{doi:10.1103/PhysRevC.21.534}}.

\bibitem{Wisshak06}
K.~{Wisshak}, F.~{Voss}, F.~{K{\"a}ppeler}, L.~{Kazakov}, {Stellar neutron
  capture cross sections of the Lu isotopes}, Phys. Rev. C 73~(1) (2006)
  015807.
\newblock \href {https://doi.org/10.1103/PhysRevC.73.015807}
  {\path{doi:10.1103/PhysRevC.73.015807}}.

\bibitem{Dillmann09}
I.~{Dillmann}, C.~{Domingo-Pardo}, M.~{Heil}, F.~{K{\"a}ppeler}, A.~{Wallner},
  O.~{Forstner}, R.~{Golser}, W.~{Kutschera}, A.~{Priller}, P.~{Steier},
  A.~{Mengoni}, R.~{Gallino}, M.~{Paul}, C.~{Vockenhuber}, {Determination of
  the stellar (n,{\ensuremath{\gamma}}) cross section of Ca40 with accelerator
  mass spectrometry}, Phys. Rev. C 79~(6) (2009) 065805.
\newblock \href {http://arxiv.org/abs/0907.0107} {\path{arXiv:0907.0107}},
  \href {https://doi.org/10.1103/PhysRevC.79.065805}
  {\path{doi:10.1103/PhysRevC.79.065805}}.

\bibitem{Macias20}
M.~{Mac{\'\i}as}, B.~{Fern{\'a}ndez}, J.~{Praena}, {The first neutron
  time-of-flight line in Spain: Commissioning and new data for the definition
  of a neutron standard field}, Radiation Physics and Chemistry 168 (2020)
  108538.
\newblock \href {https://doi.org/10.1016/j.radphyschem.2019.108538}
  {\path{doi:10.1016/j.radphyschem.2019.108538}}.

\bibitem{Ratynski88-PRC}
W.~{Ratynski}, F.~{K{\"a}ppeler}, {Neutron capture cross section of $^{197}$Au:
  A standard for stellar nucleosynthesis}, Phys. Rev. C 37~(2) (1988) 595--604.
\newblock \href {https://doi.org/10.1103/PhysRevC.37.595}
  {\path{doi:10.1103/PhysRevC.37.595}}.

\bibitem{MARTINHERNANDEZ12}
G.~Martín-Hernández, P.~Mastinu, J.~Praena, N.~Dzysiuk, R.~{Capote Noy},
  M.~Pignatari,
  \href{https://www.sciencedirect.com/science/article/pii/S0969804312003119}{Temperature-tuned
  maxwell–boltzmann neutron spectra for kt ranging from 30 up to 50kev for
  nuclear astrophysics studies}, Applied Radiation and Isotopes 70~(8) (2012)
  1583--1589.
\newblock \href
  {https://doi.org/https://doi.org/10.1016/j.apradiso.2012.05.004}
  {\path{doi:https://doi.org/10.1016/j.apradiso.2012.05.004}}.
\newline\urlprefix\url{https://www.sciencedirect.com/science/article/pii/S0969804312003119}

\bibitem{Tessler16}
M.~{Tessler}, M.~{Friedman}, S.~{Schmidt}, A.~{Shor}, D.~{Berkovits},
  D.~{Cohen}, G.~{Feinberg}, S.~{Fiebiger}, A.~{Kr{\'a}sa}, M.~{Paul},
  R.~{Plag}, A.~{Plompen}, R.~{Reifarth}, {Neutron Energy Spectra and Yields
  from the $^{7}$Li(p,n) Reaction for Nuclear Astrophysics}, in: Journal of
  Physics Conference Series, Vol. 665 of Journal of Physics Conference Series,
  2016, p. 012027.
\newblock \href {https://doi.org/10.1088/1742-6596/665/1/012027}
  {\path{doi:10.1088/1742-6596/665/1/012027}}.

\bibitem{Kappeler11}
F.~{K{\"a}ppeler}, R.~{Gallino}, S.~{Bisterzo}, W.~{Aoki}, {The s process:
  Nuclear physics, stellar models, and observations}, Reviews of Modern Physics
  83~(1) (2011) 157--194.
\newblock \href {http://arxiv.org/abs/1012.5218} {\path{arXiv:1012.5218}},
  \href {https://doi.org/10.1103/RevModPhys.83.157}
  {\path{doi:10.1103/RevModPhys.83.157}}.

\bibitem{Heil08}
M.~{Heil}, F.~{K{\"a}ppeler}, E.~{Uberseder}, R.~{Gallino}, M.~{Pignatari},
  {Neutron capture cross sections for the weak s process in massive stars},
  Phys. Rev. C 77~(1) (2008) 015808.
\newblock \href {https://doi.org/10.1103/PhysRevC.77.015808}
  {\path{doi:10.1103/PhysRevC.77.015808}}.

\bibitem{Wisshak06-PRC}
K.~{Wisshak}, F.~{Voss}, F.~{K{\"a}ppeler}, M.~{Krti{\v{c}}ka}, S.~{Raman},
  A.~{Mengoni}, R.~{Gallino}, {Stellar neutron capture cross section of the
  unstable s-process branching point $^{151}$Sm}, Phys. Rev. C 73~(1) (2006)
  015802.
\newblock \href {https://doi.org/10.1103/PhysRevC.73.015802}
  {\path{doi:10.1103/PhysRevC.73.015802}}.

\bibitem{Jarosik22}
J.~{Jaro{\v{s}}{\'\i}k}, V.~{Wagner}, M.~{Majerle}, P.~{Chudoba},
  N.~{Burianov{\'a}}, M.~{{\v{S}}tef{\'a}nik}, {Activation cross-section
  measurement of fast neutron-induced reactions in Al, Au, Bi, Co, F, Na, and
  Y}, Nuclear Instruments and Methods in Physics Research B 511 (2022) 64--74.
\newblock \href {https://doi.org/10.1016/j.nimb.2021.10.018}
  {\path{doi:10.1016/j.nimb.2021.10.018}}.

\bibitem{Vrzalova23}
J.~{Vrzalov{\'a}}, A.~{Kr{\'a}sa}, P.~{Chudoba}, J.~{Khushvaktov}, A.~{Kugler},
  M.~{Majerle}, M.~{Suchop{\'a}r}, O.~{Svoboda}, P.~{Tich{\'y}}, V.~{Wagner},
  {Excitation functions of neutron-induced threshold reactions in Au, Bi, Ta
  measured using 30-94 MeV quasi mono-energetic neutron sources}, Nucl. Phys. A
  1031 (2023) 122593.
\newblock \href {https://doi.org/10.1016/j.nuclphysa.2022.122593}
  {\path{doi:10.1016/j.nuclphysa.2022.122593}}.

\bibitem{Martin16-PRC}
G.~{Mart{\'\i}n-Hern{\'a}ndez}, P.~{Mastinu}, M.~{Maggiore}, L.~{Pranovi},
  G.~{Prete}, J.~{Praena}, R.~{Capote-Noy}, F.~{Gramegna}, A.~{Lombardi},
  L.~{Maran}, C.~{Scian}, E.~{Munaron}, {Excitation function shape and neutron
  spectrum of the $^{7}$Li(p ,n )$^{7}$Be reaction near threshold}, Phys. Rev.
  C 94~(3) (2016) 034620.
\newblock \href {https://doi.org/10.1103/PhysRevC.94.034620}
  {\path{doi:10.1103/PhysRevC.94.034620}}.

\bibitem{Macklin58-PR}
R.~L. {Macklin}, J.~H. {Gibbons}, {Study of the T(p, n)He$^{3}$ and Li$^{7}$(p,
  n)Be$^{7}$ Reactions}, Physical Review 109~(1) (1958) 105--109.
\newblock \href {https://doi.org/10.1103/PhysRev.109.105}
  {\path{doi:10.1103/PhysRev.109.105}}.

\bibitem{Taschek48-PR}
R.~{Taschek}, A.~{Hemmendinger}, {Reaction Constants for
  Li$^{7}$(p,n)Be$^{7}$}, Physical Review 74~(4) (1948) 373--385.
\newblock \href {https://doi.org/10.1103/PhysRev.74.373}
  {\path{doi:10.1103/PhysRev.74.373}}.

\bibitem{Newson57-PR}
H.~W. {Newson}, R.~M. {Williamson}, K.~W. {Jones}, J.~H. {Gibbons},
  H.~{Marshak}, {Li$^{7}$(p, n), (p, p'{\ensuremath{\gamma}}), and (p,
  {\ensuremath{\gamma}}) Reactions near Neutron Threshold}, Physical Review
  108~(5) (1957) 1294--1300.
\newblock \href {https://doi.org/10.1103/PhysRev.108.1294}
  {\path{doi:10.1103/PhysRev.108.1294}}.

\bibitem{Sekharan76-NIM}
K.~K. {Sekharan}, H.~{Laumer}, B.~D. {Kern}, F.~{Gabbard}, {A neutron detector
  for measurement of total neutron production cross sections}, Nuclear
  Instruments and Methods 133~(2) (1976) 253--257.
\newblock \href {https://doi.org/10.1016/0029-554X(76)90617-0}
  {\path{doi:10.1016/0029-554X(76)90617-0}}.

\bibitem{Gibbons59-PR}
J.~H. {Gibbons}, R.~L. {Macklin}, {Total Neutron Yields from Light Elements
  under Proton and Alpha Bombardment}, Physical Review 114~(2) (1959) 571--580.
\newblock \href {https://doi.org/10.1007/BF01832091}
  {\path{doi:10.1007/BF01832091}}.

\bibitem{Poppe76-PRC}
C.~H. {Poppe}, J.~D. {Anderson}, J.~C. {Davis}, S.~M. {Grimes}, C.~{Wong},
  {Cross sections for the $^{7}$Li(p,n)$^{7}$Be reaction between 4.2 and 26
  MeV}, Phys. Rev. C 14~(2) (1976) 438--445.
\newblock \href {https://doi.org/10.1103/PhysRevC.14.438}
  {\path{doi:10.1103/PhysRevC.14.438}}.

\bibitem{Borchers63-PR}
R.~R. {Borchers}, C.~H. {Poppe}, {Neutrons from Proton Bombardment of Lithium},
  Physical Review 129~(6) (1963) 2679--2683.
\newblock \href {https://doi.org/10.1103/PhysRev.129.2679}
  {\path{doi:10.1103/PhysRev.129.2679}}.

\bibitem{Hisatake60-JPS}
K.~{Hisatake}, Y.~{Ishizaki}, A.~{Isoya}, T.~{Nakamura}, Y.~{Nakano},
  B.~{Saheki}, Y.~{Saji}, K.~{Yuasa}, {The Reactions Li$^{7}$(p, n)Be$^{7}$,
  B$^{11}$(p, n)C$^{11}$ and Al$^{27}$(p, n)Si$^{27}$ at 8 to 14 MeV}, Journal
  of the Physical Society of Japan 15~(5) (1960) 741--748.
\newblock \href {https://doi.org/10.1143/JPSJ.15.741}
  {\path{doi:10.1143/JPSJ.15.741}}.

\bibitem{Tilley02}
D.~R. {Tilley}, C.~M. {Cheves}, J.~L. {Godwin}, G.~M. {Hale}, H.~M. {Hofmann},
  J.~H. {Kelley}, C.~G. {Sheu}, H.~R. {Weller}, {Energy levels of light nuclei
  /A=5, 6, 7}, Nucl. Phys. A 708~(1) (2002) 3--163.
\newblock \href {https://doi.org/10.1016/S0375-9474(02)00597-3}
  {\path{doi:10.1016/S0375-9474(02)00597-3}}.

\bibitem{Tilley04}
D.~R. {Tilley}, J.~H. {Kelley}, J.~L. {Godwin}, D.~J. {Millener}, J.~E.
  {Purcell}, C.~G. {Sheu}, H.~R. {Weller}, {Energy levels of light nuclei
  A=8,9,10}, Nucl. Phys. A 745~(3) (2004) 155--362.
\newblock \href {https://doi.org/10.1016/j.nuclphysa.2004.09.059}
  {\path{doi:10.1016/j.nuclphysa.2004.09.059}}.

\bibitem{Matej2022-NIMA}
Z.~{Mat{\v{e}}j}, M.~{Ko{\v{s}}{\v{t}}{\'a}l}, M.~{Majerle}, M.~{Ansorge},
  E.~{Losa}, M.~{Zme{\v{s}}kal}, M.~{Schulc}, J.~{{\v{S}}imon},
  M.~{{\v{S}}tef{\'a}nik}, J.~{Nov{\'a}k}, D.~{Koliadko}, F.~{Cvachovec},
  F.~{Mravec}, V.~{P{\v{r}}enosil}, V.~{Zach}, T.~{Czakoj}, V.~{Rypar},
  R.~{Capote}, {The methodology for validation of cross sections in quasi
  monoenergetic neutron field}, Nuclear Instruments and Methods in Physics
  Research A 1040 (2022) 167075.
\newblock \href {https://doi.org/10.1016/j.nima.2022.167075}
  {\path{doi:10.1016/j.nima.2022.167075}}.

\bibitem{Simeckova21-NPA}
E.~{{\v{S}}ime{\v{c}}kov{\'a}}, M.~{Majerle}, M.~{{\v{S}}tef{\'a}nik},
  J.~{Mr{\'a}zek}, J.~{Nov{\'a}k}, T.~{Magna}, {The activation cross section
  measurements of proton-induced reactions on Li and Ta in the energy region
  12.5-34 MeV}, Nucl. Phys. A 1016 (2021) 122310.
\newblock \href {https://doi.org/10.1016/j.nuclphysa.2021.122310}
  {\path{doi:10.1016/j.nuclphysa.2021.122310}}.

\bibitem{Majerle20-EPJ}
M.~{Majerle}, A.~V. {Prokofiev}, M.~{Ansorge}, P.~{B{\'e}m}, D.~{Hlad{\'\i}k},
  J.~{Mr{\'a}zek}, J.~{Nov{\'a}k}, E.~{{\v{S}}ime{\v{c}}kov{\'a}},
  M.~{{\v{S}}tef{\'a}nik}, {Peak neutron production from the $^{7}$Li(p,n)
  reaction in the 20-35 MeV range}, in: European Physical Journal Web of
  Conferences, Vol. 239 of European Physical Journal Web of Conferences, 2020,
  p. 20010.
\newblock \href {https://doi.org/10.1051/epjconf/202023920010}
  {\path{doi:10.1051/epjconf/202023920010}}.

\bibitem{Majerle16-NPA}
M.~{Majerle}, P.~{B{\'e}m}, J.~{Nov{\'a}k}, E.~{{\v{S}}ime{\v{c}}kov{\'a}},
  M.~{{\v{S}}tef{\'a}nik}, {Au, Bi, Co and Nb cross-section measured by
  quasimonoenergetic neutrons from p + $^{7}$Li reaction in the energy range of
  18-36 MeV}, Nucl. Phys. A 953 (2016) 139--157.
\newblock \href {https://doi.org/10.1016/j.nuclphysa.2016.04.036}
  {\path{doi:10.1016/j.nuclphysa.2016.04.036}}.

\bibitem{Ward82-PRC}
T.~E. Ward, C.~C. Foster, G.~E. Walker, J.~Rapaport, C.~A. Goulding,
  \href{https://link.aps.org/doi/10.1103/PhysRevC.25.762}{$\frac{1}{E}$
  dependence of the $^{7}\mathrm{Li}$($p$,$n$)$^{7}\mathrm{Be}$(g.s.+ 0.43 mev)
  total reaction cross section}, Phys. Rev. C 25 (1982) 762--769.
\newblock \href {https://doi.org/10.1103/PhysRevC.25.762}
  {\path{doi:10.1103/PhysRevC.25.762}}.
\newline\urlprefix\url{https://link.aps.org/doi/10.1103/PhysRevC.25.762}

\bibitem{Kalinin57}
S.~P. {Kalinin}, A.~A. {Ogloblin}, Y.~M. {Petrov}, {Excitation functions for
  the reactions Li7(p. n)Be7, B10(p.a)Be7, and B11(p. n)C11, and energy levels
  in Be8, C11, and C12 nuclei}, The Soviet Journal of Atomic Energy 2~(2)
  (1957) 193--196.
\newblock \href {https://doi.org/10.1103/PhysRev.114.571}
  {\path{doi:10.1103/PhysRev.114.571}}.

\bibitem{Generalov17_BRASP}
L.~N. {Generalov}, S.~N. {Abramovich}, S.~M. {Selyankina}, {Activation
  measurements of the integral cross sections of reactions 7Li(p, n 0 + n 1)7Be
  g.s. , 6Li(d, n 0 + n 1)7Be g.s. , 7Li(d, 2n)7Be g.s. , 65Cu(p, n)65Zn,
  65Cu(d, 2n)65Zn, and 63Cu(d, g)65Zn}, Bulletin of the Russian Academy of
  Sciences, Physics 81~(6) (2017) 644--657.
\newblock \href {https://doi.org/10.3103/S1062873817060107}
  {\path{doi:10.3103/S1062873817060107}}.

\bibitem{SRIM10}
J.~F. {Ziegler}, M.~D. {Ziegler}, J.~P. {Biersack}, {SRIM - The stopping and
  range of ions in matter (2010)}, Nuclear Instruments and Methods in Physics
  Research B 268~(11-12) (2010) 1818--1823.
\newblock \href {https://doi.org/10.1016/j.nimb.2010.02.091}
  {\path{doi:10.1016/j.nimb.2010.02.091}}.

\bibitem{Exfor14}
N.~{Otuka}, E.~{Dupont}, V.~{Semkova}, B.~{Pritychenko}, A.~I. {Blokhin},
  M.~{Aikawa}, S.~{Babykina}, M.~{Bossant}, G.~{Chen}, S.~{Dunaeva}, R.~A.
  {Forrest}, T.~{Fukahori}, N.~{Furutachi}, S.~{Ganesan}, Z.~{Ge}, O.~O.
  {Gritzay}, M.~{Herman}, S.~{Hlava{\v{c}}}, K.~{Kat{\={o}}}, B.~{Lalremruata},
  Y.~O. {Lee}, A.~{Makinaga}, K.~{Matsumoto}, M.~{Mikhaylyukova},
  G.~{Pikulina}, V.~G. {Pronyaev}, A.~{Saxena}, O.~{Schwerer}, S.~P. {Simakov},
  N.~{Soppera}, R.~{Suzuki}, S.~{Tak{\'a}cs}, X.~{Tao}, S.~{Taova},
  F.~{T{\'a}rk{\'a}nyi}, V.~V. {Varlamov}, J.~{Wang}, S.~C. {Yang},
  V.~{Zerkin}, Y.~{Zhuang}, {Towards a More Complete and Accurate Experimental
  Nuclear Reaction Data Library (EXFOR): International Collaboration Between
  Nuclear Reaction Data Centres (NRDC)}, Nuclear Data Sheets 120 (2014)
  272--276.
\newblock \href {https://doi.org/10.1016/j.nds.2014.07.065}
  {\path{doi:10.1016/j.nds.2014.07.065}}.

\bibitem{Gilmore08}
G.~Gilmore, Practical gamma-ray spectroscopy, John Wiley \& Sons, 2008.

\end{thebibliography}
\end{document}